\documentclass[10pt, twocolumn]{article}

\usepackage[utf8]{inputenc}
\usepackage{microtype} 
\usepackage{mathpazo} 
\usepackage{titlesec} 
\usepackage{xcolor} 
\usepackage{geometry}
\usepackage{fancyhdr}
\usepackage{lettrine} 
\usepackage{amsmath}
\usepackage{amssymb}
\usepackage{graphicx}
\usepackage[colorlinks=true, linkcolor=cyan!80!black, urlcolor=cyan!80!black]{hyperref}

\renewcommand{\headrulewidth}{0.5pt}
\renewcommand{\headrule}{\hbox to\headwidth{\color{black!30}\leaders\hrule height \headrulewidth\hfill}}
\titleformat{\section}{\large\bfseries\scshape\color{black!90}}{\thesection.}{0.5em}{}
\titleformat{\subsection}{\normalsize\bfseries\color{black!80}}{\thesubsection.}{0.5em}{}

\title{\vspace{-1.5cm}\huge\bfseries The Planetary Cost of AI Acceleration, Part II \\ \vspace{0.2cm} \Large The 10th Planetary Boundary and the 6.5-Year Countdown}
\author{\scshape William Zhu \, \, \, \scshape Lei Zhu}
\date{\vspace{-0.5cm}} 

\begin{document}

\maketitle
\begin{abstract}
\noindent In Part I \cite{zhu2026planetary}, we established that the planetary-level scaling of artificial intelligence (AI) can be viewed as the evolution of a thermodynamic dissipative structure, just like human civilization itself, constrained by the Earth's finite heat capacity. Building upon this framework, we incorporate empirical constraints and analyze driving forces across the levels of hardware, infrastructure, production, and ecology. 

Most crucially, the recent super-exponential scaling of autonomous Large Language Model (LLM) agents signals a \textbf{broader, fundamental paradigm shift from machines primarily replacing the human hands }(manual labor and mechanical processing)  to machines\textbf{ delegating for the human minds (cognition, reasoning, and intention)}. The uncontrolled offloading and scaling of "thinking" itself, beyond human's limited but efficient biological capacity, has profound consequences for humanity's heat balance sheet, since thinking, or intelligence, carries thermodynamic weight, as we discussed in Part I.

The Earth has already surpassed the heat dissipation threshold required for long-term ecological stability, and projecting based on empirical data reveal a concerning trajectory: without radical structural intervention, anthropogenic heat accumulation will breach critical planetary ecological thresholds in less than 6.5 years, even under the most ideal scenario where Earth Energy Imbalance (EEI) holds constant. In this work, we identify six factors from artificial intelligence that influence the global heat dissipation rate and delineate how their interplay drives society toward one of four broad macroscopic trajectories.

\textbf{We propose that the integration of artificial intelligence and its heat dissipation into the planetary system constitute the tenth planetary boundary (9+1) \cite{rockstrom2009safe}}. The core empirical measurement of this boundary is the net-new waste heat generated by exponential AI growth, balanced against its impact on reducing economic and societal inefficiencies and thus baseline anthropogenic waste heat emissions, as well as its effect on reducing the greenhouse effect. We demonstrate that managing AI scaling lacks a moderate middle ground: it will either accelerate the breach of critical planetary thermodynamic thresholds, or it will serve as the single most effective lever on stabilizing the other nine planetary boundaries and through which safeguarding human civilization's survival.

\end{abstract}
\vspace{0.5cm}

\section{New Scaling Law: Offloading Thinking Itself to the Machines}

\lettrine[lines=2, lhang=0.33, nindent=0em]{I}{n} Part I of this series \cite{zhu2026planetary}, we made projections with the assumption of there being \textbf{a sufficiently long time window for the economy to react and manage its allocation of the thermal dividends from AI’s optimization of economic inefficiencies.} However, recent developments in the AI industry point to a fundamental shift in the scaling rate. Just as the mechanization of physical labor triggered unprecedented surges in human demand, the current transition toward bulk offloading of cognitive reasoning—finally enabled by autonomous AI agents—is precipitating an explosion in the societal scale of information processing. And scaling thoughts, as they are not bound by physical constraints, is much more rapid and frictionless compared to scaling in physical labor. This is what is behind the growing month-over-month token consumption observed since early 2026, forcing computing demand onto a super-exponential curve. As a reaction to this demand, capital-driven expansion in computing infrastructure is soon expected to bring the computing hardware market to surpass trillions of dollars annually. The scale and unpredictability of the scaling rate ultimately forces a confrontation with the elephant in the room, one underlined in our previous work–\textbf{computation and intelligence are inherently heat-dissipating processes, and therefore scaling compute inextricably scales heat dissipation}.

\begin{figure}[h]
    \centering
    \includegraphics[width=0.5\textwidth]{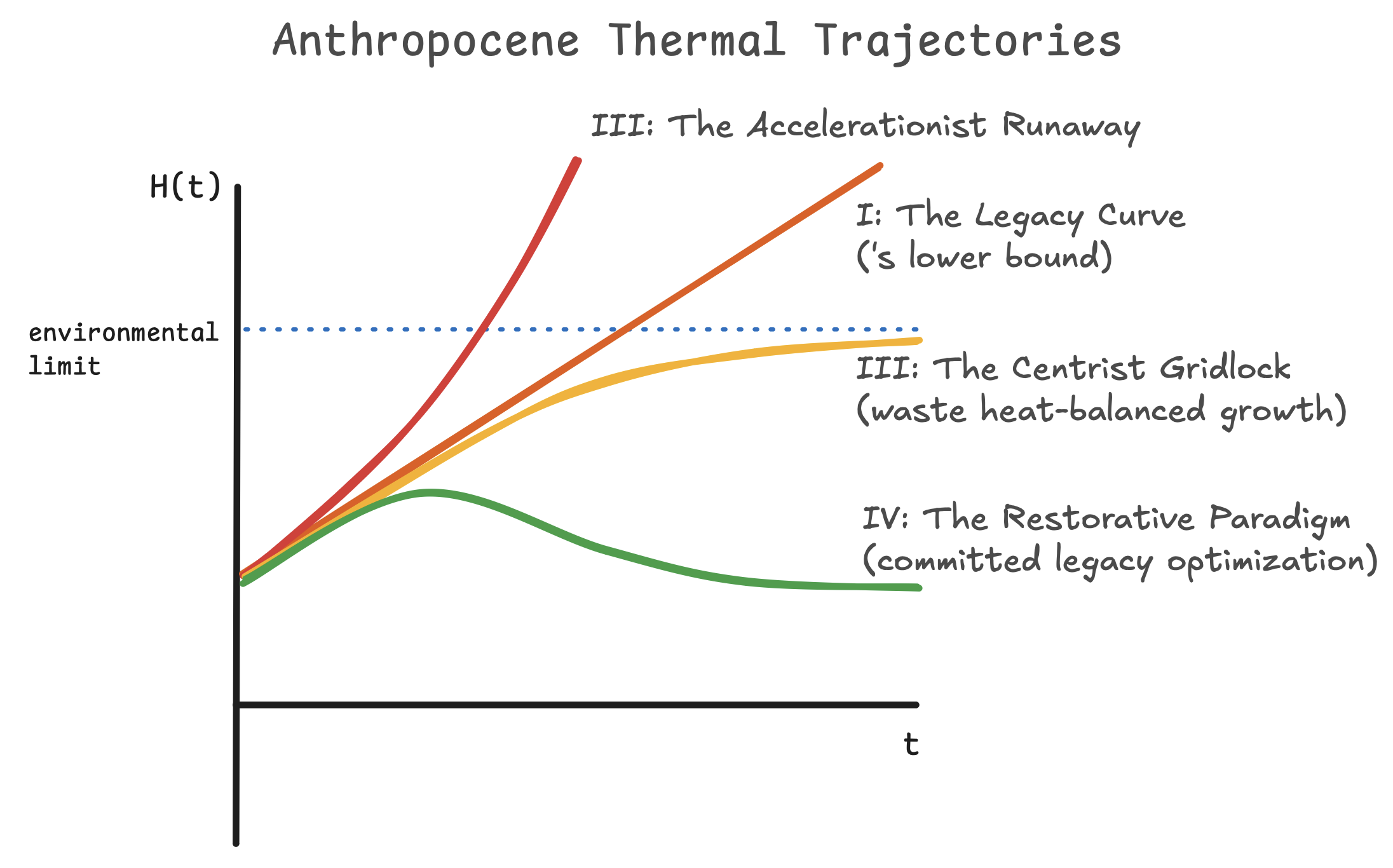}
    \caption{Net planetary heat accumulation, and four civilizational trajectories as a function of time.}
    \label{fig:heat_accumulation}
\end{figure}

\section{Reducing Heat Dissipation: Theoretical and Practical Bounds on Computation}

A prevailing techno-optimist illusion posits that future hardware efficiencies or new computing mediums (e.g., neuromorphic, photonic, or quantum architectures) will indefinitely offset this super-exponential algorithmic growth. While the Part I to this series utilized Landauer's Principle to establish the thermodynamic cost of computation (approximately $2.87 \times 10^{-21}$ Joules/bit at room temperature), this theoretical limit strictly applies at an idealized, slow setting \cite{landauer1961irreversibility}. While engineering continuously attempts to approach this physical lower bound on efficiency, useful, real-world computation operates at a speed far above this boundary, and state-of-the-art logical gate operations still generate waste heat at orders of magnitude—typically $10^5$ times \cite{markov2014limits}—above this theoretical minimum.

On the other hand, alternative hardware architectures fail to offer a reprieve. Quantum computing, for instance, requires dilution refrigerators operating near 15 millikelvins to maintain coherence. Per Carnot's Theorem, pumping microscopic heat from $0.015\text{K}$ to a $300\text{K}$ terrestrial environment necessitates massive macroscopic energy consumption, rendering its application within the Earth's ecosphere thermally futile. While relocating computing facilities to low Earth orbit is frequently proposed as a long-term solution, the inefficient thermal radiation properties of the vacuum of space—combined with thermodynamically costly launch and deployment cycles—preclude meaningful scalability for alleviating the planetary thermal load. Compounding this, massive thermal exhaust is generated during the manufacturing of any computing hardware and their heavy infrastructure. \textbf{Ultimately, we cannot navigate out of the thermodynamic tight spot we are in by only looking at hardware innovation.}

\section{Earth Energy Imbalance and the 6.5-Year Critical Period}

The primary heat reservoir on Earth is the ocean's upper mixed layer, which determines the effective climate heat capacity ($C_{eff}$) on human timescales. According to historical data from the World Climate Research Programme (WCRP), the Earth's empirical $C_{eff}$ is approximately $4.76 \times 10^{23} \text{ Joules/}^\circ\text{C}$ \cite{vonschuckmann2020heat}. Currently, humanity is only about $0.3^\circ\text{C}$ away from the globally recognized $1.5^\circ\text{C}$ critical tipping point—a threshold that, once breached, threatens to trigger an irreversible cascade of ecological collapse \cite{meng2023domino}. This leaves humanity with a finite safety buffer:
\begin{equation}
0.3^\circ\text{C} \times 4.76 \times 10^{23} \text{ Joules/}^\circ\text{C} \approx 1.42 \times 10^{23} \text{ Joules}
\end{equation}

Consequently, the velocity at which we approach this limit is dictated by the Earth Energy Imbalance (EEI). Currently, the EEI sits at a record $1.36\text{ W/m}^2$, translating to a staggering accumulation of approximately $2.19 \times 10^{22}$ Joules of un-emitted waste heat annually \cite{hansen2023global}. Crucially, this velocity is not static; it is actively accelerating. According to a landmark 2023 study by former NASA climatology head James Hansen \cite{hansen2023global} and the latest empirical data from NASA's CERES satellites, this steepening curve is driven by the dual effects of abruptly diminished aerosol cooling (such as the 2020 IMO global sulfur cap on shipping) and the continuous accumulation of greenhouse gases. Regardless, even in a highly idealized scenario where we somehow freeze the EEI at its current level indefinitely, we are left with an alarmingly brief time window before the planetary boundary is breached:
\begin{equation}
\frac{1.42 \times 10^{23} \text{ Joules}}{2.19 \times 10^{22} \text{ Joules/year}} \approx 6.5 \text{ Years}
\end{equation}
This is a time upperbound on the “legacy” human curve – natural state without AI’s involvement in economy and society. 

EEI cannot be reduced solely through the angle of reducing carbon emissions. While transitioning to a zero-carbon power grid is necessary, such a shift can only delay—rather than avert—this climate threshold. One reason is the inertia of existing atmospheric greenhouse gases, which will continue to drive EEI long after emissions cease. And more importantly, the timeline required for the large-scale deployment of new infrastructure far exceeds the projected 6.5-year window. 

While carbon emissions remain the primary driver of EEI today, the proliferation of AI introduces a significant demand-side contribution: the emission of direct waste heat. Although currently a minor factor, AI-driven heat dissipation is projected to become an increasingly substantial component of the global thermal load. Traditionally, computing served as \textbf{a utility to operate societal and economic infrastructure}—acting as \textbf{a tool or a medium for human intention and thoughts} (e.g., financial infrastructure, mobile network, or most recently social network). However, the emergence of autonomous agents shifts this role toward the direct delegation of human thought processes. Effectively, this digital amplification of 'thought' drastically increases per-capita computing demand, thereby escalating both energy consumption and thermal dissipation inherent in the algorithmic manipulation of bits, as examined earlier and in Part I of the series.

\section{Six Interacting Determinants of the Anthropocene Thermal Trajectory}

In Part I, we proposed macro-trajectories based on idealized assumptions of instantaneous AI optimization dividends, immediate realization of policies, and a dissipation rate threshold $\dot{E}_{Limit}$ yet to be surpassed. However, when confronted with the 6.5-year temporal constraint, the actual heat trajectory is determined by the interaction of six factors:

\begin{itemize}
    \item \textbf{Human Computing Demand Surge:} As agentic AI transitions from replacing manual labor to delegating cognitive processes, humans are empowered to utilize AI to “think” more than what’s previously defined by their biological limits. The accessibility of silicon-based intelligence for cognitive offloading creates a massive surge in baseline inference volume.
    \item \textbf{AI Delegation’s Recursive Growth:} As more autonomy is granted to AI agents in the next few years to solve more complex tasks and manage bigger, more abstract scopes with less human supervision, self-referential cycles emerge. In the case growth of compute budget is not strictly controlled, this leads to arbitrarily deep and increasingly opaque and unmonitored thinking traces with the agent itself or with other agents it orchestrates or collaborates with. This recursively drives a super-exponential compute demand curve independent of direct human input.
    \item \textbf{Hardware Efficiency Asymptotes:} This represents the absolute and practical floor of hardware efficiency discussed in Section 2, which defines the upper bound of computing efficiency.
    \item \textbf{Global Grid Ceiling and Infrastructure Scaling Bottleneck:} Although algorithms can iterate within weeks, the material reality of mineral extraction, hardware manufacturing, and expanding the global power supply is constrained by physical reality – we only have so much existing energy capacity, factory, and infrastructure to build new energy plants and new computing infrastructure. This imposes limits on the growth rate of AI computing power.
    \item \textbf{Economic and Societal Optimization Gains ($\dot{E}_{opt}$):} AI structurally eliminates systemic energy waste by streamlining inefficient human logistics and collaborative processes. However, this introduces a Jevons Paradox: driven by the pressures of capital expansion, global markets will inevitably leverage these efficiency gains to scale investment and consumption, ultimately generating a net increase in waste heat.
    \item \textbf{Absolute Planetary Thermodynamic Boundary:} The final, non-negotiable ecological ceiling defined by the remaining $1.42 \times 10^{23}$ Joules of permissible thermal addition.
\end{itemize}

\section{The Thermodynamic Calculus of Planetary Survival}

The question of survival shifts to a straightforward calculus problem. The thermodynamic ledger of the Anthropocene is governed by the continuous integral of the planetary EEI rate over time $t$:
\begin{equation}
\text{EEI}(t) = \dot{E}_{Legacy}(t) + \dot{E}_{AI}(t) - \dot{E}_{opt}(t)
\end{equation}

In this equation, the macroscopic dimension of $\text{EEI}(t)$ on the left side is defined as the total annual net heat flux of the entire Earth system [Joules/year]; on the right side, $\dot{E}_{Legacy}(t)$ is the baseline human waste heat emission, $\dot{E}_{AI}(t)$ is the additional computational waste heat, and $\dot{E}_{opt}(t)$ is the reduction in human waste heat emission structurally optimized by AI.

To prevent the collapse of Earth's ecology, the cumulative integral of the net heat flux must never exceed the remaining buffer:
\begin{equation}
H(t) = \int_{0}^{t} \text{EEI}(\tau) \, d\tau < 1.42 \times 10^{23} \text{ Joules}
\end{equation}

Crucially, AI energy demand ($\dot{E}_{AI}$) scales in "software time" (months), while rebuilding physical supply chains ($\dot{E}_{opt}$) takes decades. \textbf{Slowing heat accumulation is insufficient. To restore our planetary safety margin, the function $\text{EEI}(t)$ must ultimately be forced into negative territory.}

\begin{figure}[h]
    \centering
    \includegraphics[width=0.5\textwidth]{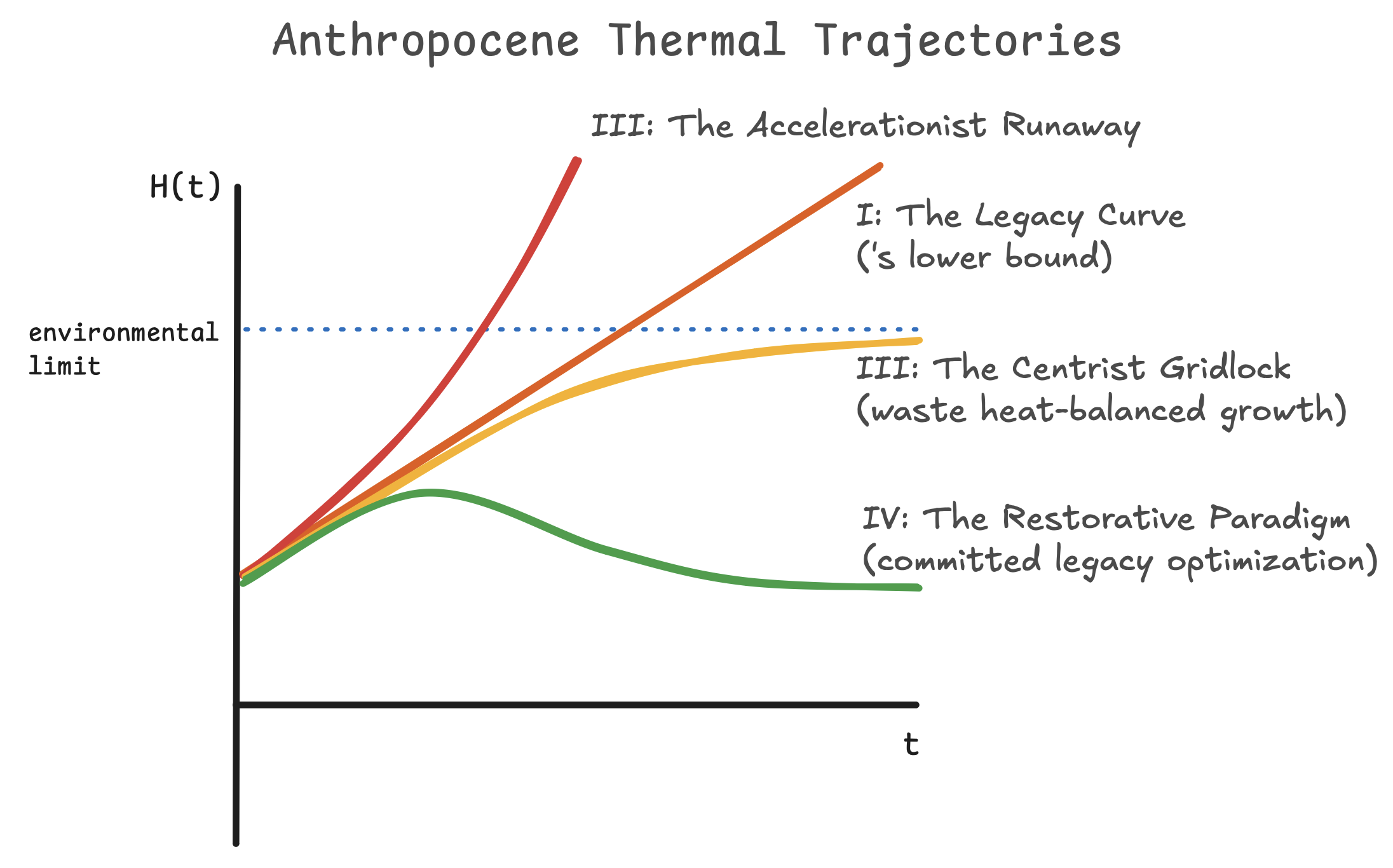}
    \caption{Net planetary heat accumulation, and four civilizational trajectories as a function of time.}
    \label{fig:heat_accumulation}
\end{figure}

\begin{figure}[h]
    \centering
    \includegraphics[width=0.5\textwidth]{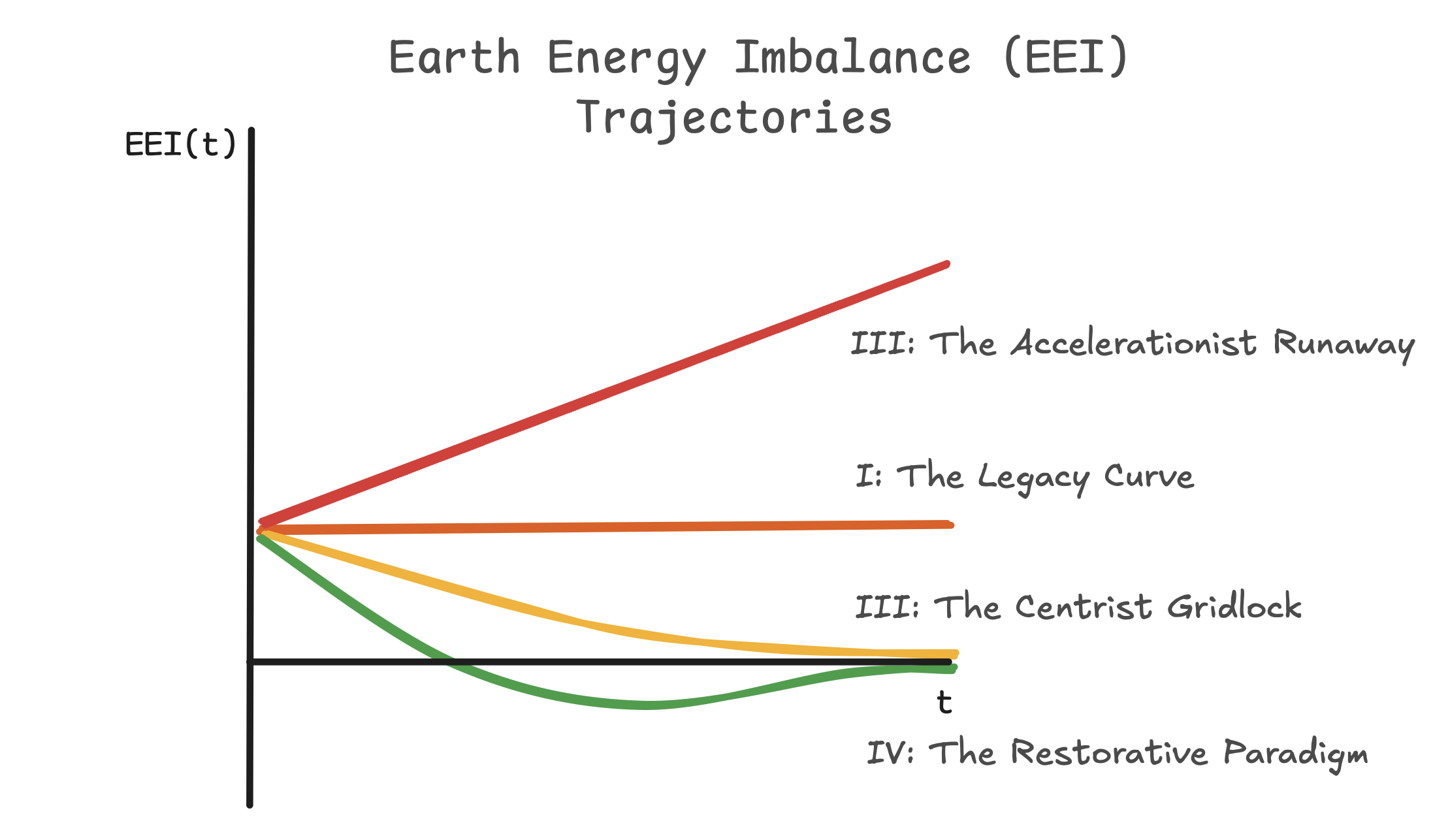}
    \caption{Earth energy imbalance (EEI), and four civilizational trajectories as a function of time}
    \label{fig:eei_time}
\end{figure}

\section{\textbf{Four Macroscopic Projections for the AI-Climate Nexus}}

Accounting for all the factors discussed above presents us with four trajectories \cite{dixson2022earth}:

\begin{enumerate}
    \item \textbf{Legacy Baseline (A Lower Bound Projection with a 6.5-Year Countdown):} This trajectory assumes a continued reliance on an unoptimized form of our economy and human collaboration ($\dot{E}_{opt} \approx 0$) but without further increase in overall EEI. In this scenario, EEI remains static at approximately $2.19 \times 10^{22} \text{ Joules/year}$. The thermal integral accumulates linearly, reaching the critical threshold in approximately 6.5 years. This triggers a catastrophic, irreversible collapse of the global ecosystem as the planetary heat capacity is exceeded.
    \item \textbf{Accelerationist Runaway (A Compressed Timeline of Collapse):} Driven by unmonitored growth in computing demand initiated by both human and AI systems themselves, the massive injection of computing waste heat ($\dot{E}_{AI}$) overwhelms any optimization gains in the short run, possibly burning through the 6.5-year buffer in an even shorter 4 to 5 years. This timeline eerily mirrors common predictions on AGI’s timeline.
    \item \textbf{Centrist Gridlock (A Zero-Margin Asymptote):} In this scenario, explosive AI growth collides with the structural supply chain limits of the global power grid. The shortage of energy headroom forces a "zero-sum" environment: to secure power quotas for further AI development, human capital or AI themselves must aggressively retrofit obsolete industrial and economic systems at every scale. Here, the reduction in waste heat from structural optimization roughly offsets the addition of computational heat ($\dot{E}_{opt} \approx \dot{E}_{AI}$). The rate of thermal integration slows, forming an asymptotic curve that approaches—but does not breach—the limit. While human civilization survives, the ecosystem remains in a state of chronic fragility, having nearly exhausted the $0.3^\circ\text{C}$ safety buffer.
    \item \textbf{Restorative Paradigm (The Only Viable Option):} The Restorative Paradigm represents the primary objective of this framework. Here, AI expansion is strictly governed by physical and thermodynamic realities. Expanding upon the strategies of Trajectory III, computational output is actively and exclusively directed toward the precise remediation of humanity's legacy heat-dissipating structures ($\dot{E}_{opt} \gg \dot{E}_{AI}$). In addition, the expanding AI capacity must also address the greenhouse effect, which contribute to heat accumulation independent of new waste heat released by the civilization, through possibly developing new approaches and technologies. By applying AI towards reducing greenhouse effects and optimizing economic and collaborative structures at all levels of humanity, this can push the annual EEI to reach zero before the cumulative sum hits the fatal threshold. Ultimately, by actively managing planetary heat dissipation, the annual $\text{EEI}(t)$ is driven to negative values until the integral generates $-1.42 \times 10^{23} \text{ Joules}$ to restore the situation at least back to where we are today.
\end{enumerate}

\section{Conclusion: AI Heat Dissipation as the Tenth Planetary Boundary}

In light of these thermodynamic imperatives, \textbf{we propose that the integration of artificial intelligence and its heat dissipation into the planetary system constitute the tenth planetary boundary \cite{rockstrom2009safe}.} The core metric of this supplementary boundary is the net-new waste heat generated by exponential AI growth balanced against its systemic impact on reducing baseline anthropogenic heat emissions. This framework stipulates that the integral of net heat flux must not breach the absolute ecological threshold of $1.42 \times 10^{23}$ Joules in the upcoming years, and that net heat flux itself must be quickly driven into a negative state. As AI represents not just a passive metric but an active lever to alleviate the pressure on the other planetary boundaries, it should be more appropriately formulated as the most important dimension in a '9+1' planetary boundary model.

Without the rehabilitative role of AI, the current projection leads to an ecosystemic collapse within 6.5 years (Trajectory I). If allowed to scale unconstrained, AI will only serve as a catalyst and accelerate Earth's ecological dissolution (Trajectory II). Under the assumption of humans having the ability to dramatically alter the annual heat accumulation rate, not only through reducing its own waste heat production from legacy systemic friction, but also reducing the effect of greenhouse effect through new technology, the thermal dividends from AI development can function as a pivotal lever to mitigate the situation.

\textbf{The definitive KPI for the AI industry should no longer be restricted to the milestones in reaching AGI, but rather the degree to which a restorative paradigm for Earth's ecosystem is realized.} Success requires alignment at every level of human civilization to utilize planetary-scale intelligence for driving a negative rate of change in thermal dissipation, repaying the historical heat debt, and ensuring our evolution toward a Kardashev Type I civilization remains safe and sound.

We are in the most critical window of phase transition that dictates the fate of our biosphere—the clock is ticking.

\bibliographystyle{unsrt}
\bibliography{main}

\end{document}